\documentclass[superscriptaddress,twocolumn,showpacs,prl]{revtex4}
\newcommand{\figurewidth}{\columnwidth}
\usepackage{graphicx,amsmath}

\begin{document}

\title{
First order phase transition in the Quantum Adiabatic
Algorithm}

\author{A.~P.~Young}
%\email{peter@physics.ucsc.edu}
\affiliation{Department of Physics, University of California,
Santa Cruz, California 95064}

\author{S.~Knysh}
\affiliation{ELORET Corporation, NASA Ames Research Center, MS 229,
Moffett Field, CA 94035-1000}
%\email{sergey.i.knysh@nasa.gov}

\author{V.~N.~Smelyanskiy}
\affiliation{NASA Ames Research Center, MS 269-3, Moffett Field, CA
94035-1000}
%\email{Vadim.N.Smelyanskiy@nasa.gov}

\date{\today}

\begin{abstract}
We simulate the quantum adiabatic algorithm (QAA) for the exact cover problem
for sizes up to $N=256$ using quantum Monte Carlo simulations incorporating
parallel tempering. At large $N$ we find that some instances have a
discontinuous (first order) quantum phase transition during the evolution of
the QAA. This fraction increases with increasing $N$ and may tend to 1 for
$N \to\infty$.

\end{abstract}
\pacs{03.67.Lx , 03.67.Ac, 64.70.Tg,75.10.Nr} \maketitle

It is of great interest to know if an eventual quantum computer could solve a
broad range of hard ``optimization'' problems more efficiently than a
classical computer.  An important class is the NP-hard category
\cite{Garey:97}, for which it is believed that all
classical algorithms take a time which grows
exponentially with the problem size $N$.

The most promising approach to solving optimization problems on a quantum
computer seems to be the
quantum adiabatic algorithm (QAA), proposed by Farhi et
al.~\cite{farhi:01}.
The idea, which is related to quantum
annealing~\cite{kadawoki:98}, is that one
adds to a ``problem" Hamiltonian, $\mathcal{H}_{\rm P}$, whose
ground state represents a solution of a classical optimization problem,
a non-commuting ``driver" Hamiltonian, $\mathcal{H}_{\rm D}$, so
the total Hamiltonian is
\begin{equation}
\mathcal{H}(s) = (1-s) \mathcal{H}_{\rm D} +
s \mathcal{H}_{\rm P}, \label{qu_ham}
\end{equation}
where $s(t)$ is a \textit{time dependent} control
parameter.
$\mathcal{H}_{\rm P}$
is expressed
in terms of classical Ising spins taking values $\pm 1$, or equivalently
in terms of the $z$-components of
the Pauli matrices for each spin, $\hat\sigma^z_i$. The simplest driver
Hamiltonian is then then $\mathcal{H}_{\rm D} = -\sum_{i=1}^N
\hat\sigma^x_i$ where $\hat\sigma^x_i$ is the $x$-component Pauli matrix.

The control parameter $s(t)$ is 0 at $t=0$, so
$\mathcal{H}$=$\mathcal{H}_{\rm D}$, which has a trivial ground state in
which all $2^N$ basis states (in the $\hat\sigma^z$ basis) have equal
amplitude. It then increases with $t$, reaching 1 at $t={\cal T}$, where
${\cal T}$ is the runtime of the algorithm, at which
point $\mathcal{H}$=$\mathcal{H}_{\rm P}$. If the time evolution of
$s(t)$ is sufficiently slow, the process will be adiabatic. Hence,
starting the system in the ground state of $\mathcal{H}_{\rm D}$,
the system will end up in the ground
state of $\mathcal{H}_{\rm P}$ and the problem is solved.
The time
$\mathcal{T}$ required to find the ground state with significant probability
is called the complexity.
The bottleneck of the QAA is likely to be
at one or more points where the
energy gap from the ground state to the first excited state becomes
very small, possibly due to a quantum phase transition.

Early numerical  work~\cite{farhi:01,hogg:03}
on very small systems, $N \le 24$
(for a particular constraint
satisfaction problem known as ``exact cover 3'', also called
1-in-3 sat),
found that the complexity scaled
polynomially with size, roughly as $N^2$,
which caused a good deal of excitement.
However, this power law complexity
may be an artifact of the very small sizes studied, so it is of great
interest to determine whether the complexity continues to be polynomial
for much larger sizes or whether a ``crossover'' to exponential
complexity is seen.  

In previous work~\cite{young:08} we have used quantum Monte Carlo (QMC)
simulations to investigate much larger
sizes of the exact cover problem, up to $N = 128$.
%\textit{sampling} of the states is is performed.
We found evidence that, while the median complexity is still polynomial, 
an increasing fraction of instances became very hard to
equilibrate for the larger sizes.
We have
now considerably improved the algorithm, borrowing techniques from the
the spin glass field to speed up equilibration. We have therefore been
able to understand much better
these ``troublesome" instances, and find that they have a first order quantum
phase transition. Furthermore, we have increased the range of sizes still
further, up to $N = 256$, finding that the fraction instances with a
first order transition continues to increase with $N$, plausibly tending
1 for $N \to\infty$. 
The gap at a 
first order phase transition is likely to be exponentially
small~\cite{altshuler:09,amin:09}, and hence
lead to exponential complexity for the QAA.

We now describe the model and our results in more detail.
To make a comparison with the earlier work we study (essentially) the
same model for $\mathcal{H}_{\rm P}$ used by Farhi et
al.~\cite{farhi:01}. This problem, known as exact cover, 
is a \textit{random satisfiability} problem, a class which is known to be  
NP-hard. In exact cover there are $N$ Ising
spins  and $M$ ``clauses'' each of which involves three spins (chosen at
random). The energy of a clause is zero if one spin is $-1$ and the other
two are $1$, and otherwise the energy is a positive integer. The
simplest Hamiltonian with this property is~\cite{different}
\begin{equation}
\mathcal{H}_{\rm P} = {1 \over 4}
\sum_{\alpha=1}^M 
\, \left(\hat\sigma^z_{\alpha_1} +
\hat\sigma^z_{\alpha_2} + \hat\sigma^z_{\alpha_3} - 1\right)^2 , 
\label{hclass}
\end{equation}
where $\alpha_1, \alpha_2$ and $\alpha_3$ are the three spins in clause
$\alpha$, and the $\{\hat\sigma^{z}_i\} (i=1,\cdots,N)$ are Pauli matrices.
In the absence of the driver Hamiltonian, the Pauli matrices can be
replaced by classical Ising spins $S_i$ taking values $\pm 1$.
An
instance has a ``satisfying assignment'' if there is at least one choice
for the spins where the total energy is zero. As the ratio $\alpha
\equiv M/N$ is
increased, there is a phase transition at $\alpha_s$ where the number of satisfying
assignments goes to zero.  The version used by Farhi et al.~considers
only instances with a \textit{unique} satisfying assignment (USA), i.e.
there is only \textit{one} state with energy 0. This has the advantage
that the gap $\Delta E(s)$ between the ground state and first excited
state is greater than zero in both limiting cases,
${\cal H}=\mathcal{H}_{\rm D}$ and ${\cal H}=\mathcal{H}_{\rm P}$, but
will have a \textit{minimum} at an intermediate value $s = s^*$. In
addition, it ensures that we work close to the satisfiability transition
where the problem is particularly hard~\cite{kirkpatrick:94}.
Hence here, and in the earlier work~\cite{young:08},
we consider instances with a USA.

The method of generating instances with a USA is described in Ref.~\cite{young:08}.
For each size $N$ we choose the number of clauses $M$ which maximizes the
probability of finding a USA, see Table \ref{tab:M}. The actual number of
spins simulated $N'$, is somewhat less than $N$ due to
isolated sites being omitted, and others that
do not affect the complexity are also ``pruned off''~\cite{young:08}.
The value of $\alpha \equiv M/N$ seems to be close to the critical value
$\alpha_s \simeq 0.626$~\cite{raymond:07} for $N \to \infty$.

\begin{table}
\begin{center}
\begin{tabular}{|r|r|r|r|r|r|r|r|}
\hline\hline
N        & 16     & 32     & 64     & 128    & 192    & 256 \\
\hline
M        & 12     & 23     & 44     & 86     & 126    & 166 \\
\hline
$\alpha$ & 0.7500 & 0.7188 & 0.6875 & 0.6719 & 0.6563 & 0.6484 \\
\hline\hline
\end{tabular}
\end{center}
\caption{The sizes studied in the simulation.
}
\vspace{-0.5cm}
\label{tab:M}
\end{table}

In QMC 
we simulate an effective classical model with Ising spins
$S_i(\tau)=\pm1$ in which $\tau$ ($0 \le \tau < \beta \equiv
T^{-1}$) is imaginary time.  Following common practice, we
discretize imaginary time 
into $L_\tau$ ``time slices'' each representing $\Delta \tau =
\beta/L_\tau$ of imaginary time. We take $\Delta\tau = 1$.

As discussed previously~\cite{young:08}, for $\beta\Delta E \gg 1$ (where
$\Delta E \equiv E_1 - E_0$ is the energy gap), and $\tau \ll \beta$,
the time-dependent 
correlation function
\vspace{-0.2cm}
\begin{equation}
C(\tau) = {1 \over N' L_\tau} \sum_{i=1}^{N'} \sum_{\tau_0=1}^{L_\tau}
\langle\, S_i(\tau_0 + \tau) S_i(\tau_0)\, \rangle \, ,
\end{equation}
is a sum of exponentials, i.e.
\begin{equation}
C(\tau) = q + \sum_{n\geq 1} A_n
\exp[-(E_n - E_0) \tau] \, , \label{ctau}
\end{equation}
where the $A_n$
are constants and $q$, the spin glass order parameter, is given by
\begin{equation}
q = {1 \over N'}
\sum_{i=1}^{N'} 
\langle \, S_i \, \rangle^2 \, .
\label{q}
\end{equation} 
At large $\tau$, the sum in Eq.~\eqref{ctau} is dominated
by the term corresponding to the first excited state, ($n=1$), and so
$\Delta E \equiv E_1 - E_0$ can be obtained by fitting $\log [C(\tau) - q]$ against
$\tau$ for large $\tau$.

We have considerably improved the algorithm relative to that in
Ref.~\cite{young:08} by incorporating ``parallel
tempering"~\cite{Hukushima:96,marinari:98b}, which has been very successful in
speeding up simulations of spin glass systems. Whereas in spin glasses,
one simulates copies of the system at different, close-by temperatures,
in the quantum case, the copies are at different values of the control
parameter $s$.

As already mentioned, the focus of the present
study is to determine which instances have a first order transition.
Parallel tempering is very good at equilibrating the system on either
side of the transition. However, it is still difficult (i) to determine exactly
where the transition occurs, because both phases are metastable in the
region where they are not the equilibrium state, and (ii) to accurately 
determine the minimum gap for first-order instances, because it is so
small. We have performed runs starting the spins both from a random
initial configuration, and from the solution of the problem Hamiltonian.
If we start by ``seeding'' the spins with the exact solution, we are
confident that the Monte Carlo is in the correct phase for $s$
close to 1. It is also in the correct phase for small $s$ because
equilibration is easy in this region. Hence, if a long simulation
starting the spins from the exact solution produces a sharp
discontinuity,
%i.e.~a first order transition,
we feel that this is almost
certainly the correct behavior.

\begin{figure}
\begin{center}
\includegraphics[width=\figurewidth]{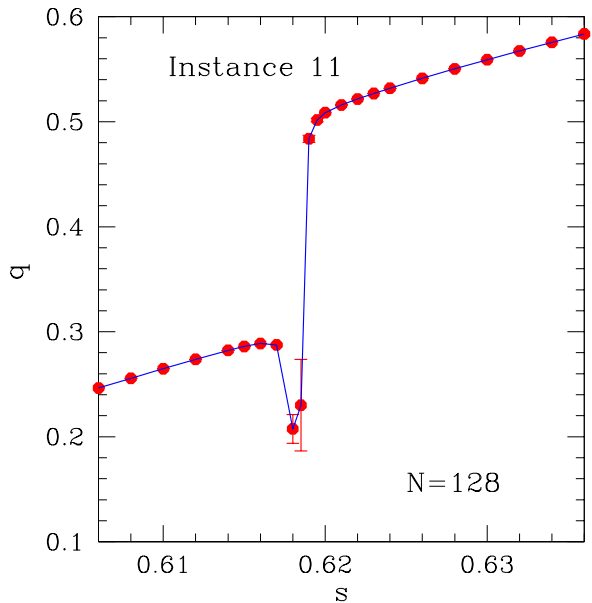}
\caption{(Color online) 
An instance with a first order transition with $N = 128$. Note
the expanded horizontal scale.
}
\label{allq_N128_11}
\vspace{-0.7cm}
\end{center}
\end{figure}

In order to investigate whether or not a first order transition occurs we
focus on the spin glass order parameter $q$ defined in Eq.~\eqref{q}.
The expectation value of $q$ is always non-zero
because of terms linear in the $\hat\sigma^z$ (magnetic field terms) in the
Hamiltonian, Eq.~\eqref{hclass}.
To
determine the square of the average without bias we simulated two copies of
the spins at each value of $s$ and evaluate $\langle S_i \rangle^2$ as
$\langle S_i \rangle^{(1)} \langle S_i \rangle^{(2)}$. A representative result
for an instance with a first order transition is shown in
Fig.~\ref{allq_N128_11}.
Note the very rapid increase in $q$ over a very small
range of $s$, and that the two curves on each side of the jump are
obviously displaced vertically with respect to each other.

The dip before the jump clearly seen
in Fig.~\ref{allq_N128_11} provides clear evidence for a
two-phase coexistence, and hence a first order transition, for the following
reason.
If both copies are in the
same phase, then the mean value of $\langle S_i\rangle$ is the same in both
copies. However, right at the first order transition, one copy can be in one
phase (the low-$q$ phase, say) and the other copy in the other (high-$q$)
phase. The average value of $\langle S_i \rangle$ can have different signs in
the two phases for
\textit{some} sites $i$.
Hence, the typical Hamming distance between the spin configurations in
the two copies can be even greater (and so $q$ even smaller) than when both
copies are in the low-$q$ phase. In every instance where we observed a sharp
jump, this was proceeded by a dip. Hence we use the dip as a
precise criterion for a
first order transition.

\begin{figure}
\begin{center}
\includegraphics[width=\figurewidth]{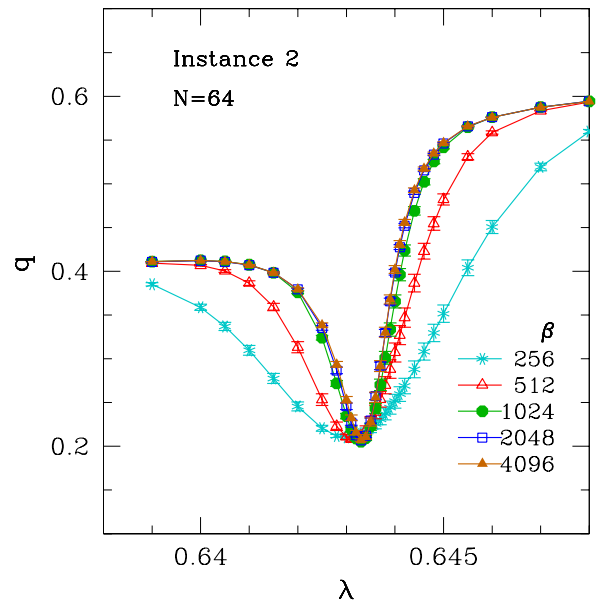}
\caption{(Color online) 
The main figure shows the spin glass order parameter $q$, defined in
Eq.~\eqref{q}, as a function of $s$
for an instance with $N = 64$ which has a first order
transition. The different curves are for different values of $\beta$. The
inset shows the energy gap $\Delta E$ as a function of $s$ for $\beta =
2048$, indicating that $\Delta E_\text{min} = 0.0021$ (same value was found
for $\beta = 1024$ and $4096$). From the main figure one
sees that the width of the finite-size rounding increases with $T \equiv
1/\beta$ for $T \gg \Delta E$ but is independent of $T$ in the opposite limit
$T \ll \Delta E$, as expected from Eq.~\eqref{fssrounding}. Note the 
expanded horizontal scale.
\label{qdip}
}
\vspace{-0.7cm}
\end{center}
\end{figure}

Of course, even a first order transition is rounded out for a finite-size
system.
To estimate the
size of the rounding
we need to consider the two cases $\Delta E_\text{min} \gg T$ and
$\Delta E_\text{min} \ll T$ separately, where $\Delta E_\text{min}$ is
the minimum value of the gap at the transition. If $\Delta E_\text{min}
\gg T$, $\delta s$ is the range of $s$ over which
$\Delta E$ changes by an amount $\Delta E_\text{min}$, whereas if
$\Delta E_\text{min}
\ll T$, $\delta s$ is the range of $s$ over which 
$\Delta E$ changes by an amount equal to $T$. Hence
\begin{equation}
\delta s = \left\{
\begin{array}{lll}
\Delta E_\text{min} \, \left(\displaystyle \partial\Delta E \over 
\displaystyle \partial s
\right)^{-1} , & (\Delta E_\text{min} \gg T) , \medskip \\
T \, \left(\displaystyle \partial\Delta E \over
\displaystyle \partial s
\right)^{-1} ,
& (\Delta E_\text{min} \ll T) . 
\end{array}
\right.
\label{fssrounding}
\end{equation}
Figure \ref{qdip} shows the finite-size rounding
for an instance
with $N=64$, small enough that we can equilibrate
through the (first order) 
transition. For $\beta \lesssim 1024$ the width of the transition region
increases as $\beta \equiv 1/T$ decreases, but for $\beta \gtrsim 1024$ the
width is independent of $\beta$. For this instance we find $\Delta E_\text{min}
= 0.0021$ as shown in the inset, so the width of the rounding
becomes independent of $T$ when $T \ll \Delta E$ as expected.

\begin{figure}
\begin{center}
\includegraphics[width=\figurewidth]{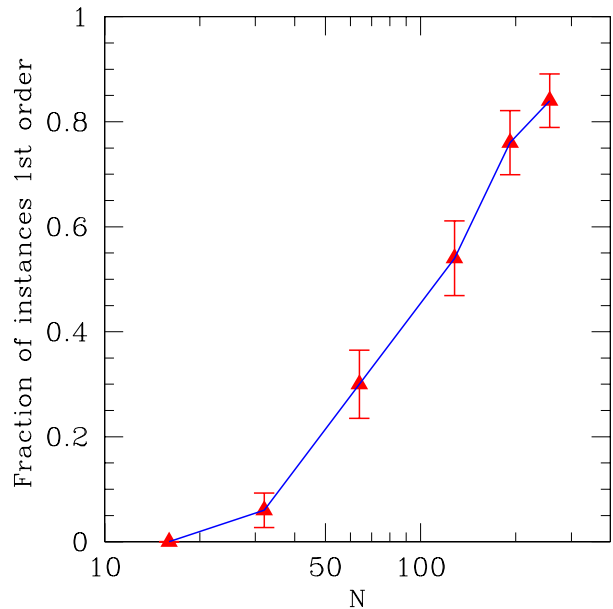}
\caption{(Color online) 
The fraction of instances with a first order transition (defined in the way
discussed in the text) as a function of size. For each size, 50
instances were studied.
\label{frac_firstorder}
}
\vspace{-0.7cm}
\end{center}
\end{figure}

In Fig.~\ref{frac_firstorder} we plot the fraction of
instances with a first order transition. For each size we have studied
$N_\text{inst}= 50$
instances. If we denote the first-order fraction by $r$ 
then the error bar in $r$ is
$\sqrt{r(1-r)/(N_\text{inst}-1)}$. The figure shows that $r$
increases rapidly with $N$ and,
very plausibly, tends to 1 for $N \to \infty$.  We see that the first order
fraction is slightly greater than a half for $N=128$. In our earlier
work~\cite{young:08} we found that the \textit{median} complexity continued to
be polynomial up to $N=128$ (the largest size studied). However, there is no
contrast with the present work because, as already noted in
Ref.~\cite{different}, the
models used are slightly different, and as a result the crossover to a first order
transition occurs at a slightly lower value of $N$ in the present model. The
crossover to first order would have been seen in the earlier model if
somewhat larger sizes had been studied.

Exponentially small gaps have been discussed before in the context of the QAA.
Some time ago, one of us~\cite{smelyanskiy:02}
pointed out for a different problem, number
partitioning,
that the minimum gap is exponentially small, because of a transition between
the states that are "localized" and "extended" in the computational basis.

Altshuler et al.~\cite{altshuler:09} predict an
exponentially small gap at large $N$ for exact cover.
Performing perturbation theory away from $s = 1$ they argue that
there will be a level crossing between two ``localized'' states
for $s$ close to 1 at which point the
ground state configuration changes abruptly.
In our numerics, there is a big variation in
the location of the first order transition for a given size, but we do not
detect a systematic shift towards $s = 1$ as the size increases. However,
Altshuler et al.~predict that $1 -s \sim N^{-1/8}$, which is
probably too slow to be visible
in our data.
It will be interesting to investigate in future work whether the first order
transition found here is due to the mechanism they propose.

Farhi et al.~\cite{farhi:09} used a continuous
imaginary time QMC method to study a very similar problem to ours,
except that two solutions far away in Hamming space
are ``planted'' into the Hamiltonian. This ensures that there is a finite 
probability of a
first order transition where the equilibrium state changes from one
planted solution to another.  By contrast, our work does nothing explicit to
impose a first order transition.

J\"org et al.\cite{jorg:08} studied quantum annealing for 
the quantum random energy model (REM), the classical version of which~\cite{REM}
has a ``1-step
replica symmetry breaking" (also called a ``random first order'') transition.
Following Goldschmidt~\cite{goldschmidt:90},
they find a discontinuous quantum transition and argue that this leads to
an exponentially small gap.
They also observed that an exponentially small gap is seen
in quantum versions of several models with random first order
transitions and suggested that this may be the general feature
of all such models, including satisfiability~\cite{monasson:99,raymond:07}.
%We note, however,
%that
However,
the classical REM
has zero spin glass order
parameter $q$ in the disordered phase~\cite{REM}
whereas classical random satisfiability
models have lower symmetry because $q$ is \textit{always} non-zero 
due to the terms linear in 
$\hat\sigma^z$ in Eq.~\eqref{hclass}.
Consequently, it is not obvious to us that the first
order quantum transition observed here is due to the same mechanism as 
that found~\cite{jorg:08} for the quantum REM. Very recently, a first order
transition has also been found in another model by J\"org et al.\cite{jorg:09}.

To conclude, we have a found a crossover to a first order quantum phase
transition during the evolution of the QAA
for instances of exact cover with a unique satisfying
assignment when the size becomes greater than about 100. It is possible that
the complexity for \textit{random}
instances of exact cover could be different.
We are therefore studying
instances of exact cover with the USA constraint removed, and will also study
other models in addition to exact cover.

\begin{acknowledgments}
We thank Eddie Farhi, Florent Krzakala, Boris Altshuler
and Mike Moore for helpful
discussions and correspondence.
The work of APY is supported in part 
by the National Security Agency (NSA) 
under Army Research Office (ARO)
contract number W911NF-09-1-0391, and by a Special Research Grant
from the Committee
on Research at UCSC.
%used to purchase computers.
%for this project.
The work of SK and VNS is supported by
National Security Agency's Laboratory of Physics Sciences and the NASA Ames
NAS Supercomputing Center. 
We are grateful to Andre Petukhov for generously allowing us a
to use the Gamow computer cluster at the South Dakota
School of Mines and Technology.
\end{acknowledgments}

\bibliography{refs,comments}

\begin{thebibliography}{20}
\bibitem[{dif()}]{different}
\bibinfo{note}{This is slightly different from the Hamiltonian used in
  Ref.~[\onlinecite{young:08}] in which a 3-spin term is added so that the
  energy is \textit{always} 1 if the sum of the spins is different from 1. This
  explains some minor differences with Ref.~[\onlinecite{young:08}].}
\end{thebibliography}

\end{document}